# Symmetry, Conserved Charges, and Lax Representations of Nonlinear Field Equations: A Unified Approach

C. J. Papachristou

*Department of Physical Sciences, Naval Academy of Greece, Piraeus 18539, Greece*

E-mail:  papachristou@snd.edu.gr

**Abstract:** A certain non-Noetherian connection between symmetry and integrability properties of nonlinear field equations in conservation-law form is studied. It is shown that the symmetry condition alone may lead, in a rather straightforward way, to the construction of a Lax pair, a doubly infinite set of (generally nonlocal) conservation laws, and a recursion operator for symmetries. Applications include the chiral field equation and the self-dual Yang-Mills equation.

## 1. Introduction

In a recent paper [1] an analytical method was described for constructing a Lax pair for the Ernst equation of General Relativity. The starting point was the symmetry condition (or linearized form) of the field equation. The latter equation is in conservation-law form, and thus so is its associated symmetry condition. A doubly infinite hierarchy of conservation laws was then constructed by a recursive process, and the conserved "charges" were used as Laurent coefficients in a series representation (in powers of the spectral parameter) of a function $\Psi$ which was seen to satisfy the sought-for Lax pair.

It is natural to inquire whether this technique can also be applied to other nonlinear partial differential equations (PDEs) of Mathematical Physics. This article describes a general, non-Noetherian framework for connecting integrability characteristics of a given nonlinear PDE to the symmetry properties of this PDE. It is remarkable that, by starting with the symmetry condition, one may discover a number of important things such as the existence of a recursion operator [2,3] for symmetries, a doubly-infinite set of (typically nonlocal) conservation laws, and a Lax pair which "linearizes" the nonlinear field equation.

To illustrate the use of the method, application is made to two familiar nonlinear PDEs: the chiral field equation and the self-dual Yang-Mills equation. In these examples, the corresponding Lax pairs and infinite sequences of conservation laws are constructed explicitly. Moreover, the recursion operators for symmetries are derived. In the case of the real Ernst equation, treated previously in [1], although a recursion operator doesn't seem to exist for that particular form of the equation (due to the coordinate "pathology" which results in the explicit appearance of an independent variable in the PDE), one still gets an interesting "hidden" symmetry transformation which leads to new approximate solutions for stationary gravitational fields with axial symmetry [4].

## 2.  The General Idea

Let $F[u]=0$ be a nonlinear PDE in the dependent variable $u$ and the independent variables $x, y, \ldots$ . The bracket notation $[u]$ indicates that the function $F$ may depend explicitly on the variables $u, x, y, \ldots$ , as well as on partial derivatives, of various orders, of $u$ with respect to the independent variables, denoted $u_x$, $u_y$, $u_{xx}$, $u_{yy}$, $u_{xy}$, etc. We adopt the definition according to which the PDE is integrable if it has an associated Lax-pair representation, i.e., if it can be expressed as an integrability condition for solution of a linear system of PDEs for an auxiliary field $\Psi$ :

$$L_i(\Psi; u; \lambda) = 0 \quad i=1,2 \qquad (1)$$

where the differential expressions $L_i$ are linear in $\Psi$, and where $\lambda$ is a (generally complex) "spectral" parameter.

It has been observed that integrable PDEs often have an infinite number of symmetries which may be produced, for example, with the aid of one or more recursion operators (see, e.g., [2,3] and the references therein). This connection between symmetry and integrability may be attributed to a variety of factors. For example, an integrable PDE may have an underlying Hamiltonian structure in which the Lagrangian density possesses an infinite number of variational symmetries. In this case, the Noether theorem provides the connection between symmetry and integrability, the latter manifesting itself in the presence of an infinite set of conservation laws. As is often the case, the existence of these laws is associated with a Lax structure for the nonlinear problem.

Non-Noetherian connections between symmetry and integrability, however, may also exist. Let us recall that the nonlinear PDE $F[u]=0$ is a consistency condition for solution of the system (1). On the other hand, the (generally complex) function $\Psi$ will satisfy some PDE of its own, also derived from system (1). This PDE will be linear in $\Psi$ and will contain $u$ as a "parametric" function:

$$G(\Psi; u) = 0 \qquad (2)$$

where the expression $G$ is linear in $\Psi$, and where $u$ is a solution of $F[u]=0$ . We may say that the system (1) is a Bäcklund transformation relating the nonlinear PDE $F[u]=0$ to the linear PDE (2). Now, we already know an equation of the form (2): it is the *symmetry condition* (linearized form) of $F[u]=0$ . Let $u' = u + \alpha Q[u]$ be an infinitesimal symmetry transformation for the latter PDE, where $\alpha$ is an infinitesimal parameter (we note that any symmetry of a PDE can be expressed as a transformation of the dependent variable alone [2,3], i.e., is equivalent to a "vertical" symmetry). The *symmetry characteristic* $Q[u]$ then satisfies a linear PDE of the form

$$S(Q; u) = 0 \quad \mathrm{mod}\, F[u] \qquad (3)$$

where "mod $F[u]$" signifies that the PDE on the left is satisfied when $u$ is a solution of the nonlinear PDE $F[u]=0$ . Now, if it happens that Eqs.(2) and (3) become identical when $\Psi \equiv Q$ (i.e., if the functions $G$ and $S$ are the same), then the solution $\Psi$ of the Lax pair (1) will also be a symmetry characteristic of $F[u]=0$ :

$$S(\Psi; u) = 0 \quad \mathrm{mod}\, F[u] \qquad (4)$$

Of course, as the examples of the self-dual Yang-Mills equation [5] and the Ernst equation [1,4] have taught us, it is possible that a given nonlinear PDE admit more than one Lax representation. What we are seeking here is a Lax pair which functions as a Bäcklund transformation connecting the nonlinear PDE $F[u]=0$ to its (linear) symmetry condition (3). The symmetry condition itself is thus "built" into the Lax pair, and a very fundamental connection between symmetry and integrability is established.

With regard to the complex parameter $\lambda$ of the Lax pair (1), we remark the following: Since the role of such a parameter is generally nontrivial, it will be required that $\lambda$ be nonzero (as well as, of course, finite in magnitude). We then expect that the solution $\Psi$ of system (1), for a given $u$ satisfying $F[u]=0$, will be an analytic function of $\lambda$ for $\lambda \neq 0$. This solution may thus be represented as a Laurent series expansion in powers of $\lambda$, with $u$-dependent coefficients:

$$\Psi(u;\lambda) = \sum_{n=-\infty}^{+\infty} \lambda^n Q^{(n)}[u] \qquad (5)$$

where the functions $Q^{(n)}[u]$ may be local or nonlocal in $u$. Now, we recall that $\Psi$ is assumed to be a symmetry characteristic of $F[u]=0$, and this must be true for all values of $\lambda$ in the Lax pair. Substituting Eq.(5) into Eq.(4) (which is linear in $\Psi$), and equating coefficients of all $\lambda^n$ to zero, we find a doubly infinite set of linear PDEs for $Q^{(n)}$, of the form

$$S(Q^{(n)};u) = 0 \mod F[u], \quad n = 0, \pm 1, \pm 2, \cdots \qquad (6)$$

All Laurent coefficients $Q^{(n)}[u]$ are thus seen to be symmetry characteristics for the nonlinear PDE $F[u]=0$, and the presence of this infinite set of symmetries is intimately related to the Lax pair.

Substituting the expansion (5) into the Lax pair (1), and equating coefficients of all powers of $\lambda$ to zero, we obtain a pair of linear PDEs containing $Q^{(n)}$ and (say) $Q^{(n+1)}$. In essence, this is a Bäcklund transformation for the symmetry condition (3). This differential recursion relation constitutes a *recursion operator* [2,3] for the PDE $F[u]=0$, in the spirit of a new perception of this concept originally proposed by this author [5,6] and, independently, by Marvan [7]. We thus have a method for the explicit construction of such an operator. Starting with any symmetry $Q^{(0)}$, we can, in principle, use this operator to derive a double infinity of symmetries $Q^{(n)}$ (although not all of them will necessarily be nontrivial).

Finally, suppose that $F[u]$ is a divergence, so that the PDE $F[u]=0$ has the form of a conservation law. Then, its symmetry condition (3) also is in such form. Given that an infinite number of symmetry characteristics $Q^{(n)}[u]$ are available, we immediately obtain a doubly infinite collection of conservation laws for $F[u]=0$ from Eq.(6) (where now the function $S$ is a divergence). Typically, the recursion operator connecting the $Q^{(n)}$ to each other is an integro-differential operator; thus, the conserved "currents" are generally expected to be nonlocal in $u$.

## 3. Analytical Description of the Method

Our objective is the following: Given a nonlinear PDE $F[u]=0$ in conservation-law form, we seek a Lax pair whose solution is a symmetry characteristic for this PDE, and, in the process, we expect to derive a recursion operator for symmetries as well as an infinite set of (nonlocal) conservation laws. Although the solution $u$ of the PDE may depend on more than two independent variables, we restrict ourselves to the case where $F[u]$ is a divergence in only two of them:

$$F[u] \equiv D_x A[u] + D_y B[u] = 0 \qquad (7)$$

where $D_x$ and $D_y$ denote total derivatives (see Appendix), which will also be indicated by using subscripts: $D_x A \equiv A_x$, etc. We will assume, in general, that $u$ is square-matrix-valued, and so are the functions $A$, $B$, $F$.

Let $\delta u = \alpha Q[u]$ be an infinitesimal symmetry of Eq.(7) (where $\alpha$ is an infinitesimal parameter and $Q$ is the matrix-valued symmetry characteristic). We write, in finite form,

$$\Delta u = Q[u] \qquad (8)$$

where, in general, $\Delta$ denotes the Fréchet derivative of any function $f[u]$, with respect to the characteristic $Q$ (see Appendix). The symmetry condition for the PDE (7) is

$$\Delta F[u] = 0 \quad \mod F[u] \qquad (9)$$

where

$$\Delta F[u] = D_x \Delta A[u] + D_y \Delta B[u]$$

(since Fréchet derivatives and total derivatives commute). Putting

$$\Delta A[u] \equiv G(Q;u), \quad \Delta B[u] \equiv H(Q;u) \qquad (10)$$

(where the functions $G$ and $H$ are linear in $Q$), we rewrite Eq.(9) in the form of a linear PDE for $Q$:

$$S(Q;u) \equiv D_x G(Q;u) + D_y H(Q;u) = 0 \quad \mod F[u] \qquad (11)$$

We note that $S(Q\,;u)$ is a divergence, so that the symmetry condition (11) is a conservation law for the corresponding nonlinear PDE (7).

Equation (11) suggests that we introduce a "potential" function $K$, such that $G=K_y$ and $H=-K_x$ (subscripts denote total differentiations). We assume that $K$ is linearly dependent on some new function $Q'$, and we write:

$$G(Q;u) = D_y K(Q';u), \quad H(Q;u) = -D_x K(Q';u) \qquad (12)$$

Clearly, this system is integrable for $Q'$ (mod $F[u]$) if $Q$ satisfies the symmetry condition (11). The integrability requirement for $Q$, on the other hand, will yield some linear PDE for $Q'$. It is possible that, by an appropriate choice of the function $K(Q';u)$, this PDE will be just the symmetry condition (11) for $Q'$:

$$S(Q';u) = 0 \mod F[u] .$$

That is, $Q'$ will also be a symmetry characteristic. The system (12) then constitutes a Bäcklund transformation (BT) for the symmetry condition (11). This BT may be viewed as an invertible recursion operator for symmetries of the nonlinear PDE (7). Such an operator will, in principle, produce a doubly infinite sequence of symmetry characteristics $Q^{(n)}$ ($n = \pm 1, \pm 2, ...$) from any given characteristic $Q^{(0)}$.

To better display the recursive character of the BT (12), we rewrite this system as follows:

$$G(Q^{(n)};u) = D_y K(Q^{(n+1)};u)$$
$$H(Q^{(n)};u) = -D_x K(Q^{(n+1)};u) \tag{13}$$

($n = 0, \pm 1, \pm 2, ...$), where $G$ and $H$ are linear in $Q^{(n)}$, while $K$ is linear in $Q^{(n+1)}$. Now, since all the $Q^{(n)}$ satisfy the PDE (11), the BT (13) also yields a double infinity of conservation laws for the field equation (7):

$$D_x G(Q^{(n)};u) + D_y H(Q^{(n)};u) = 0 \mod F[u] \tag{14}$$

Starting with a known symmetry characteristic $Q$, we can evaluate the conserved "charges" $Q^{(n)}$ ($n = 0, \pm 1, \pm 2, ...$) as follows: (*a*) We take the BT (13) with $n=0$ and set $Q^{(0)} = Q$ on the left-hand side. Then, $Q^{(1)}$ is found by integration. To find $Q^{(2)}$ we similarly integrate the BT (13) with $n=1$, etc. We thus obtain all positively-indexed charges $Q^{(n)}$. (*b*) We take the BT (13) with $n = -1$ and set $Q^{(0)} = Q$ on the right-hand side. We then solve for $Q^{(-1)}$. Working similarly for $n = -2, -3, ...$, we obtain all negatively-indexed charges $Q^{(n)}$.

We now introduce a complex parameter $\lambda$, which we require to be nonzero and of finite magnitude. Multiplying both sides of Eq.(13) by $\lambda^n$, summing over all integral values of $n$, and taking into account that the functions $G$, $H$ and $K$ are linear in their respective $Q$'s, we find the following pair of PDEs:

$$G\left(\sum_{n=-\infty}^{+\infty} \lambda^n Q^{(n)};u\right) = \frac{1}{\lambda} D_y K\left(\sum_{n=-\infty}^{+\infty} \lambda^n Q^{(n)};u\right)$$
$$H\left(\sum_{n=-\infty}^{+\infty} \lambda^n Q^{(n)};u\right) = -\frac{1}{\lambda} D_x K\left(\sum_{n=-\infty}^{+\infty} \lambda^n Q^{(n)};u\right) \tag{15}$$

We set

$$\Psi(u;\lambda) = \sum_{n=-\infty}^{+\infty} \lambda^n Q^{(n)}[u] \qquad (16)$$

Equation (16) has the form of a Laurent expansion of a complex function $\Psi$ in powers of $\lambda$, for a given solution $u$ of the field equation (7). We note that $\Psi$ is a linear combination of symmetry characteristics of Eq.(7), hence $\Psi$ itself is a symmetry characteristic of that PDE. Substituting Eq.(16) into Eq.(15), we rewrite the latter in the form of a system of linear PDEs for $\Psi$:

$$D_y K(\Psi;u) = \lambda G(\Psi;u) , \qquad D_x K(\Psi;u) = -\lambda H(\Psi;u) \qquad (17)$$

The consistency of this system requires that $\Psi$ satisfy the linear PDE (11),

$$S(\Psi;u) \equiv D_x G(\Psi;u) + D_y H(\Psi;u) = 0 \pmod{F[u]} .$$

This verifies that $\Psi$ is a symmetry characteristic. Moreover, the system (17) is linear in $\Psi$, and its solvability demands that $u$ satisfy the nonlinear PDE (7) [this was required from the start in order that the BT (13), by which the charges $Q^{(n)}$ appearing in the Laurent expansion (16) are defined, may be integrable for $Q^{(n)}$ and $Q^{(n+1)}$]. We thus conclude that the linear system (17) constitutes a Lax pair for the field equation (7), and that, moreover, the solution $\Psi$ of this system is a symmetry characteristic of that equation.

A final comment before closing this section: The whole idea was based on the assumption that an auto-Bäcklund transformation of the form (12) exists for the symmetry condition (11). It is possible, however, that no choice for the function $K(Q';u)$ in Eq.(12) exists such that $Q'$ be a symmetry characteristic when $Q$ is such a characteristic. In this case, the method described above may still furnish an infinite number of conservation laws as well as a Lax pair, albeit not a recursion operator for producing infinite sets of symmetries. Moreover, the solution $\Psi$ of the Lax pair will no longer represent a symmetry of the field equation (although, of course, it will somehow be related to a symmetry, since the symmetry condition was the starting point for constructing the Lax pair). The example of the Ernst equation, examined in detail in [1], made this point clear. In this case, the absence of an infinite set of symmetries is not a property of the gravitational field equations themselves (which, when properly formulated, do exhibit such an infinite set [8]) but is a consequence of the chosen real form of the Ernst equation, in which a spatial coordinate makes an explicit appearance.

## 4. Chiral Field Equation

The chiral field equation (a two-dimensional reduction of the self-dual Yang-Mills equation, to be discussed later) is of the form

$$F[g] \equiv (g^{-1} g_t)_t + (g^{-1} g_x)_x = 0 \qquad (18)$$

where $g$ is a $GL(N,C)$-valued function of $t$ and $x$ (as usual, subscripts denote total differentiations with respect to these variables). Let $\delta g = \alpha Q[g]$ be an infinitesimal symmetry of Eq.(18), with symmetry characteristic $Q[g]$. We have that $\Delta g = Q[g]$, where $\Delta$ denotes the Fréchet derivative with respect to $Q$ (see Appendix). Moreover, by the commutativity of the Fréchet derivative with total derivatives,

$$\Delta F[g] = D_t \Delta(g^{-1} g_t) + D_x \Delta(g^{-1} g_x)$$
$$= D_t \hat{A}_t (g^{-1} Q) + D_x \hat{A}_x (g^{-1} Q)$$

where we have introduced the "covariant derivative" operators

$$\hat{A}_t = D_t + [g^{-1} g_t,\ ]\ ,\qquad \hat{A}_x = D_x + [g^{-1} g_x,\ ]$$

(the square brackets denote commutators). It can be shown that these operators commute, as expected from the fact that the "connections" $g^{-1} g_t$ and $g^{-1} g_x$ are pure gauges. The symmetry condition (9) reads:

$$S(Q;g) \equiv (D_t \hat{A}_t + D_x \hat{A}_x)(g^{-1} Q) = 0 \mod F[g] \tag{19}$$

and it is obviously in conservation-law form.

We now seek an auto-Bäcklund transformation (BT) of the form (12) for the linear PDE (19). This must be of the form

$$\hat{A}_t (g^{-1} Q) = K_x\ ,\qquad \hat{A}_x (g^{-1} Q) = -K_t$$

for some function $K(Q';g)$. Let us try $K(Q';g) = g^{-1} Q'$:

$$\hat{A}_t (g^{-1} Q) = (g^{-1} Q')_x\ ,\qquad \hat{A}_x (g^{-1} Q) = -(g^{-1} Q')_t \tag{20}$$

Integrability for $Q'$ clearly requires that $Q$ satisfy Eq.(19). The integrability condition for $Q$ can be written (by taking into account that covariant derivatives commute),

$$[\hat{A}_t, \hat{A}_x](g^{-1} Q) = 0\ .$$

After a somewhat lengthy calculation, and by using the operator identity

$$\hat{A}_t D_t + \hat{A}_x D_x = D_t \hat{A}_t + D_x \hat{A}_x - [F[g],\ ]$$
$$= D_t \hat{A}_t + D_x \hat{A}_x \mod F[g]$$

we find that the above integrability condition yields the PDE

$$(D_t \hat{A}_t + D_x \hat{A}_x)(g^{-1} Q') = 0 \mod F[g]$$

which is just the symmetry condition (19) for $Q'$. We conclude that Eq.(20) is indeed an auto-BT for the aforementioned symmetry condition. This BT is equivalent to a recursion operator for symmetries of the field equation (18). It can be rewritten in the form (13), as follows:

$$\hat{A}_t(g^{-1}Q^{(n)}) = D_x(g^{-1}Q^{(n+1)})$$
$$\hat{A}_x(g^{-1}Q^{(n)}) = -D_t(g^{-1}Q^{(n+1)})$$
(21)

($n= 0, \pm1, \pm2, ...$). The conservation laws of the form (14) (which form a doubly infinite set) are written, in this case,

$$(D_t\hat{A}_t + D_x\hat{A}_x)(g^{-1}Q^{(n)}) = 0 \mod F[g]$$
(22)

(where all conserved "charges" $Q^{(n)}$ are symmetry characteristics), while the Lax pair (17) reads,

$$D_x(g^{-1}\Psi) = \lambda\,\hat{A}_t(g^{-1}\Psi)\,,\quad D_t(g^{-1}\Psi) = -\lambda\,\hat{A}_x(g^{-1}\Psi)$$
(23)

The proof of the Lax-pair property of the linear system (23) is sketched as follows: By the integrability condition $(g^{-1}\Psi)_{xt} = (g^{-1}\Psi)_{tx}$, we get:

$$S(\Psi;g) \equiv (D_t\hat{A}_t + D_x\hat{A}_x)(g^{-1}\Psi) = 0$$
(24)

On the other hand, the integrability condition $\lambda\,[\hat{A}_t, \hat{A}_x](g^{-1}\Psi) = 0$, yields:

$$S(\Psi;g) - \left[F[g], g^{-1}\Psi\right] = 0$$

which, in view of Eq.(24), becomes $\left[F[g], g^{-1}\Psi\right] = 0$. This is valid independently of $\Psi$ if $F[g]=0$, i.e., if $g$ is a solution of the field equation (18). We conclude that the linear system (23) is indeed a Lax pair for the nonlinear PDE (18), the solution $\Psi$ of which pair is a symmetry characteristic [as follows from Eq.(24)]. We note that this Lax pair is different from that found several years ago by Zakharov and Mikhailov [9].

We conclude this section by giving an example of using the BT (21) to find conserved charges $Q^{(n)}$. Let us consider the symmetry characteristic $Q^{(0)} = gM$, where $M$ is an arbitrary constant matrix. The BT (21) with $n=0$, integrated for $Q^{(1)}$, yields

$$Q^{(1)} = g[X, M]\,,$$

where $X$ is the potential of Eq.(18), defined by the system of equations

$$g^{-1}g_t = X_x\,,\quad g^{-1}g_x = -X_t$$
(25)

We note that $Q^{(1)}$ is the characteristic of a *potential symmetry* [3,6]. Higher-order charges $Q^{(n)}$ with $n>1$ (which also are higher-order potential symmetries) are similarly found by recursive integration of the BT (21) with $n=1, 2$, etc.

To find negatively-indexed charges and corresponding symmetries, we begin with the BT (21) with $n=-1$, which we integrate for $Q^{(-1)}$. The result is a rather uninteresting local symmetry: $Q^{(-1)} = \Lambda g$, where $\Lambda$ is any constant matrix. Iterating for $n=-2$, however, we find a new characteristic $Q^{(-2)}$, given by the system of equations

$$Q_t - Q g^{-1} g_t = g (g^{-1} \Lambda g)_x \; , \quad Q_x - Q g^{-1} g_x = -g (g^{-1} \Lambda g)_t$$

(where we have put $Q^{(-2)} = Q$, for brevity). Higher-order, negatively-indexed charges are obtained by further iteration.

Unfortunately, in contrast to the "internal" symmetries considered above, the local coordinate symmetries [such as $Q^{(0)} = g_t$, $Q^{(0)} = g_x$, etc.] do not yield any new results by applying the BT (21). These latter symmetries, however, play an equally important role as internal ones in problems in more than two dimensions, as the example discussed in the next section will show.

## 5. Self-Dual Yang-Mills Equation

The self-dual Yang-Mills (SDYM) equation is written in the form

$$F[J] \equiv (J^{-1} J_y)_{\bar{y}} + (J^{-1} J_z)_{\bar{z}} = 0 \qquad (26)$$

where $J$ is assumed *SL(N,C)*-valued (i.e., det $J=1$). The four independent variables (appearing as subscripts) are constructed from the coordinates of an underlying complexified Euclidean space in such a way that $\bar{y}$ and $\bar{z}$ become the complex conjugates of $y$ and $z$, respectively, when the above space is real. As usual, subscripts denote total derivatives with respect to these variables.

Let $\delta J = \alpha Q[J]$ be an infinitesimal symmetry of Eq.(26), where the characteristic $Q$ is subject to the condition that $tr(J^{-1} Q) = 0$, required for producing new *SL(N,C)* solutions from old ones. The symmetry condition is, in analogy with Eq.(19),

$$S(Q;J) \equiv (D_{\bar{y}} \hat{A}_y + D_{\bar{z}} \hat{A}_z)(J^{-1} Q) = 0 \quad \mathrm{mod}\ F[J] \qquad (27)$$

where we have introduced the covariant derivatives

$$\hat{A}_y = D_y + [J^{-1} J_y, \ ] \; , \quad \hat{A}_z = D_z + [J^{-1} J_z, \ ]$$

(note again that these operators commute). An auto-BT for the linear PDE (27) [analogous to that of Eq.(20)], which is consistent with the physical requirement $tr(J^{-1} Q) = 0$, is the following:

$$\hat{A}_y(J^{-1}Q) = (J^{-1}Q')_{\bar{z}} \ , \quad \hat{A}_z(J^{-1}Q) = -(J^{-1}Q')_{\bar{y}} \qquad (28)$$

Integrability for $Q'$ requires that $Q$ satisfy Eq.(27). Integrability for $Q$, expressed by the condition $[\hat{A}_y, \hat{A}_z](J^{-1}Q) = 0$, and upon using the operator identity

$$\hat{A}_y D_{\bar{y}} + \hat{A}_z D_{\bar{z}} = D_{\bar{y}}\hat{A}_y + D_{\bar{z}}\hat{A}_z - [F[J], \ ]$$
$$= D_{\bar{y}}\hat{A}_y + D_{\bar{z}}\hat{A}_z \ \mathrm{mod}\ F[J]$$

leads us again to Eq.(27), this time for $Q'$. The BT (28) may be regarded as an invertible recursion operator for the SDYM equation. It can be re-expressed as

$$\begin{aligned}\hat{A}_y\left(J^{-1}Q^{(n)}\right) &= D_{\bar{z}}\left(J^{-1}Q^{(n+1)}\right) \\ \hat{A}_z\left(J^{-1}Q^{(n)}\right) &= -D_{\bar{y}}\left(J^{-1}Q^{(n+1)}\right)\end{aligned} \qquad (29)$$

($n = 0, \pm 1, \pm 2, ...$). From this we get a double infinity of conservation laws of the form

$$(D_{\bar{y}}\hat{A}_y + D_{\bar{z}}\hat{A}_z)\left(J^{-1}Q^{(n)}\right) = 0 \ \mathrm{mod}\ F[J] \qquad (30)$$

Finally, the Lax pair for SDYM [analogous to those of Eqs.(17) and (23)] is

$$D_{\bar{z}}(J^{-1}\Psi) = \lambda\,\hat{A}_y(J^{-1}\Psi) \ , \quad D_{\bar{y}}(J^{-1}\Psi) = -\lambda\,\hat{A}_z(J^{-1}\Psi) \qquad (31)$$

The proof of the Lax-pair property is sketched as follows: By the integrability condition $(J^{-1}\Psi)_{\bar{z}\bar{y}} - (J^{-1}\Psi)_{\bar{y}\bar{z}} = 0$, we get:

$$S(\Psi;J) \equiv (D_{\bar{y}}\hat{A}_y + D_{\bar{z}}\hat{A}_z)(J^{-1}\Psi) = 0 \qquad (32)$$

On the other hand, the integrability condition $\lambda[\hat{A}_z, \hat{A}_y](J^{-1}\Psi) = 0$, yields:

$$S(\Psi;J) - \left[F[J], J^{-1}\Psi\right] = 0$$

which, in view of Eq.(32), becomes $\left[F[J], J^{-1}\Psi\right] = 0$. This is valid independently of $\Psi$ if $F[J]=0$, i.e., if $J$ is an SDYM solution. We conclude that the linear system (31) is indeed a Lax pair for the SDYM equation (26), the solution $\Psi$ of which pair is a symmetry characteristic [as follows from Eq.(32)]. This Lax pair can be shown to be equivalent to that reported previously by this author [5], although the systematic method for explicitly constructing this system is presented here for the first time.

We now give examples of using the BT (29) to find conserved charges $Q^{(n)}$. Let us consider the symmetry characteristic $Q^{(0)} = JM$, where $M$ is a constant traceless matrix. The BT (29) with $n=0$, integrated for $Q^{(1)}$, yields

$$Q^{(1)} = J[X, M] \ ,$$

where $X$ is the potential of Eq.(26), defined by the system of equations

$$J^{-1} J_y = X_{\bar{z}} \ , \quad J^{-1} J_z = -X_{\bar{y}} \tag{33}$$

We note that $Q^{(1)}$ is the characteristic of a potential symmetry [3,6]. Higher-order charges $Q^{(n)}$ with $n>1$ (which also are higher-order potential symmetries) are similarly found by recursive integration of the BT (29) with $n=1, 2$, etc.

To find negatively-indexed charges and corresponding symmetries, we begin with the BT (29) with $n=-1$, which we integrate for $Q^{(-1)}$. The result is a familiar local symmetry: $Q^{(-1)} = \Lambda J$, where $\Lambda$ is any constant traceless matrix. Iterating for $n=-2$, we find a new characteristic $Q^{(-2)}$, given by the system of equations

$$Q_y - Q J^{-1} J_y = J (J^{-1} \Lambda J)_{\bar{z}} \ , \quad Q_z - Q J^{-1} J_z = -J (J^{-1} \Lambda J)_{\bar{y}}$$

(where we have put $Q^{(-2)} = Q$, for brevity). Higher-order, negatively-indexed charges are obtained by further iteration.

In the preceding example, the initial symmetry characteristic $Q^{(0)}$ represented an "internal" symmetry (a symmetry in the fiber space). Local coordinate symmetries (symmetries in the base space), however, also lead to the discovery of infinite sets of potential symmetries and associated conservation laws for SDYM. As an example, consider the obvious symmetry of $y$-translation, represented by the characteristic $Q^{(0)} = J_y$. The BT (29) with $n=0$, integrated for $Q^{(1)}$, gives

$$Q^{(1)} = J X_y$$

[where $X$ is the SDYM potential defined in Eq.(33)], which is another potential symmetry. Higher-order potential symmetries, whose characteristics $Q^{(n)}$ ($n>0$) appear as conserved charges in conservation laws of the form (30), can be found by repeated application of the recursion operator (29). The infinite sets of potential symmetries generated by coordinate transformations have been shown to possess a rich Lie-algebraic structure [10,11].

To conclude our example, let us find some negatively-indexed symmetries. The BT (29) with $n=-1$ and $Q^{(0)} = J_y$, integrated for $Q^{(-1)}$, gives: $Q^{(-1)} = J_{\bar{z}}$, which is the characteristic for the obvious $\bar{z}$-translational symmetry. The first nontrivial result is found for $n=-2$, yielding a characteristic $Q^{(-2)}$ which is defined by the set of equations

$$Q_y - Q J^{-1} J_y = J (J^{-1} J_{\bar{z}})_{\bar{z}} \ , \quad Q_z - Q J^{-1} J_z = -J (J^{-1} J_{\bar{z}})_{\bar{y}}$$

(where we have put $Q^{(-2)} = Q$, for brevity).

## 6. Summary

Motivated by the results of [1] for the Ernst equation, we have proposed a general, non-Noetherian scheme for connecting symmetry and integrability properties of nonlinear PDEs in conservation-law form. We have shown that, by starting with the symmetry condition (which is itself a local conservation law for the associated nonlinear PDE), one may derive significant mathematical objects such as a recursion operator for symmetries, a Lax pair, and an infinite collection of (generally nonlocal) conservation laws. Such objects are usually sought by trial-and-error processes, thus any systematic technique for their discovery is useful.

The method was illustrated by using two physically significant examples, namely, the chiral field equation and the self-dual Yang-Mills (SDYM) equation. The latter PDE has been shown to constitute a prototype equation from which several other integrable PDEs are derived by reduction [12,13]. Thus, the results regarding SDYM may also prove useful for the study of other nonlinear problems.

## 7. Appendix: Total Derivatives and Fréchet Derivatives

To make this article as self-contained as possible, we define two key concepts that are being used, namely, the total derivative and the Fréchet derivative. The reader is referred to the extensive review article [14] by this author for more details. (It should be noted, however, that our present definition of the Fréchet derivative corresponds to the definition of the Lie derivative in that article. Since these two derivatives are *locally* indistinguishable, this discrepancy in terminology should not cause any concern mathematically.)

We consider the set of all PDEs of the form $F[u]=0$, where, for simplicity, the solutions $u$ (which may be matrix-valued) are assumed to be functions of only two variables $x$ and $t$ : $u=u(x,t)$. In general, $F[u] \equiv F(x,t,u,u_x,u_t,u_{xx},u_{tt},u_{xt},\cdots)$. Geometrically, we say that the function $F$ is defined in a *jet space* [2,15] with coordinates $x$, $t$, $u$, and as many partial derivatives of $u$ as needed for the given problem. A solution of the PDE $F[u]=0$ is then a surface in this jet space.

Let $F[u]$ be a given function in the jet space. When differentiating such a function with respect to $x$ or $t$, both implicit (through $u$) and explicit dependence of $F$ on these variables must be taken into account. If $u$ is a scalar quantity, we define the *total derivative operators* $D_x$ and $D_t$ as follows:

$$D_x = \frac{\partial}{\partial x} + u_x \frac{\partial}{\partial u} + u_{xx} \frac{\partial}{\partial u_x} + u_{xt} \frac{\partial}{\partial u_t} + \cdots$$

$$D_t = \frac{\partial}{\partial t} + u_t \frac{\partial}{\partial u} + u_{xt} \frac{\partial}{\partial u_x} + u_{tt} \frac{\partial}{\partial u_t} + \cdots$$

(note that the operators $\partial/\partial x$ and $\partial/\partial t$ concern only the explicit dependence of $F$ on $x$ and $t$). If, however, $u$ is matrix-valued, the above representation has only symbolic significance and cannot be used for actual calculations. We must therefore define the total derivatives $D_x$ and $D_t$ in more general terms.

We define a linear operator $D_x$, acting on functions $F[u]$ in the jet space and having the following properties:

1. On functions $f(x,t)$ in the base space,

$$D_x f(x,t) = \partial f / \partial x \equiv \partial_x f .$$

2. On functions $F[u] = u$ or $u_x$, $u_t$, etc., in the "fiber" space,

$$D_x u = u_x , \quad D_x u_x = u_{xx} , \quad D_x u_t = u_{tx} = u_{xt} , \quad \text{etc.}$$

3. The operator $D_x$ is a derivation on the algebra of all functions $F[u]$ in the jet space (i.e., the Leibniz rule is satisfied):

$$D_x (F[u] G[u]) = (D_x F[u]) G[u] + F[u] D_x G[u] .$$

We similarly define the operator $D_t$. Extension to higher-order total derivatives is obvious (although these latter derivatives are no longer derivations, i.e., they do not satisfy the Leibniz rule). The following notation has been used in this article:

$$D_x F[u] \equiv F_x[u] , \quad D_t F[u] \equiv F_t[u] .$$

Finally, it can be shown that, for any matrix-valued functions $A$ and $B$ in the jet space, we have

$$(A^{-1})_x = -A^{-1} A_x A^{-1} , \quad (A^{-1})_t = -A^{-1} A_t A^{-1}$$

and

$$D_x [A, B] = [A_x, B] + [A, B_x] , \quad D_t [A, B] = [A_t, B] + [A, B_t]$$

where square brackets denote commutators.

Let now $\delta u \simeq \alpha Q[u]$ be an infinitesimal symmetry transformation (with characteristic $Q[u]$) for the PDE $F[u]=0$. We define the *Fréchet derivative* with respect to the characteristic $Q$ as a linear operator $\Delta$ acting on functions $F[u]$ in the jet space and having the following properties:

1. On functions $f(x,t)$ in the base space,

$$\Delta f(x,t) = 0$$

(this is a consequence of our liberty to choose all our symmetries to be in "vertical" form [2,3]).

2. On $F[u]=u$,
$$\Delta u = Q[u] .$$

3. The operator $\Delta$ commutes with total derivative operators of any order.

4. The Leibniz rule is satisfied:

$$\Delta(F[u]\,G[u]) = (\Delta F[u])\,G[u] + F[u]\Delta G[u]\ .$$

The following properties can be proven:

$$\Delta u_x = (\Delta u)_x = Q_x[u]\ ,\quad \Delta u_t = (\Delta u)_t = Q_t[u]$$

$$\Delta(A^{-1}) = -A^{-1}(\Delta A)\,A^{-1}\ ;\quad \Delta[A,B] = [\Delta A, B] + [A, \Delta B]$$

where $A$ and $B$ are any matrix-valued functions in the jet space.

If the solution $u$ of the PDE is a scalar function (thus so is the characteristic $Q$), the Fréchet derivative with respect to $Q$ admits a differential-operator representation of the form

$$\Delta = Q\frac{\partial}{\partial u} + Q_x\frac{\partial}{\partial u_x} + Q_t\frac{\partial}{\partial u_t} + Q_{xx}\frac{\partial}{\partial u_{xx}} + Q_{tt}\frac{\partial}{\partial u_{tt}} + Q_{xt}\frac{\partial}{\partial u_{xt}} + \cdots$$

Such representations, however, are not valid for PDEs in matrix form. In these cases we must resort to the general definition of the Fréchet derivative given above.

Finally, by using the Fréchet derivative, the symmetry condition for a PDE $F[u]=0$ can be expressed as follows [2,3]:

$$\Delta F[u] = 0 \mod F[u]\ .$$

This condition yields a linear PDE for the symmetry characteristic $Q$, of the form

$$S(Q;u) = 0 \mod F[u]\ .$$

## Acknowledgment

I thank Kathleen O'Shea-Arapoglou for carefully reviewing the manuscript and making a number of useful suggestions.